\documentclass{PoS}

\usepackage{amssymb}
\usepackage{amsmath}
\usepackage{amsthm}
\usepackage{makecell}

\title{Searching for the doubly-charged Higgs bosons in the Georgi-Machacek model at the ep colliders}

\ShortTitle{Searching for the doubly-charged Higgs bosons in the GM model at the ep colliders}

\author{\speaker{Hao Sun}
        and Xuan Luo\\
        Institute of Theoretical Physics, School of Physics,\\
        Dalian University of Technology, \\
        No.2 Linggong Road, Dalian, Liaoning, 116024, P.R.China\\
        E-mail: \email{haosun@dlut.edu.cn},\ \ \ \ \email{xuanluo@mail.dlut.edu.cn}}

\abstract{
The Georgi-Machacek model is one of many beyond Standard Model scenarios
with an extended scalar sector which can group under the custodial $\rm SU(2)_C$ symmetry.
There are 5-plet, 3-plet and singlet Higgs bosons under the classification of such symmetry
in addition to the Standard Model Higgs boson.
Here we study the prospects for detecting the doubly-charged Higgs boson ($\rm H_5^{\pm\pm}$)
through the vector boson fusion production at the electron-proton colliders.
Typically, we concentrate on our analysis through $\mu$-lepton pair production via pair of same-sign W bosons decay.
The discovery significance are calculated as the functions of the triplet vacuum expectation value
and necessary luminosity.
}

\FullConference{XXV International Workshop on Deep-Inelastic Scattering and Related Subjects\\
		3-7 April 2017\\
		University of Birmingham, UK}

\begin{document}

\section{Introduction}

The 125 GeV Standard Model (SM)-like particle was observed at the Large Hadron Collider (LHC).
Even so, it may still too early to conclude it is the whole story that responsible for
the electroweak symmetry breaking and even the mass of all the elementary particles.
In fact, from a theoretical point of view, there is no fundamental reason for a minimal Higgs sector,
as occurs in the SM. It is therefore motivated to consider additional Higgs representations
that also contribute to the symmetry breaking, and by doing this, one may even hold answers
to some longstanding questions in particle physics, such as the origin of neutrino mass,
the identity of the dark matter, and may establish a relationship with a yet undiscovered sector.

In order to extend the Higgs sector in the SM, we can add isospin singlet or isospin doublet directly
to the SM Higgs doublet, or even higher isospin multiplet. No matter which way we use,
the following two requirements from the experimental data should be taken into account:
one is that the electroweak $\rho$ parameter should be very close to unity and
the other is that the tree level flavor changing neutral current processes should be strongly suppressed.
Collider experiments are dedicating their efforts to search for signal of additional Higgs particles
which arise in a number of scenarios, one of them is the Georgi-Machacek (GM) model.

Here we study the prospects for detecting the doubly-charged Higgs boson ($\rm H_5^{\pm\pm}$)
via the vector boson fusion production at the electron-proton (ep) colliders.
We concentrate on our analysis through $\mu$-lepton pair production via pair of same-sign W decay.
The talk is arranged as below: we review the Georgi-Machacek model in section 2 shortly.
Section 3 is arranged to present the numerical calculations. Finally comes the short summary in last section.

\section{The Georgi-Machacek Model}

In the GM model, two $\rm SU(2)_L$ isospin triplet scalar fields,
$\chi$ (with hypercharge $\rm Y=2$) and $\xi$ (with $\rm Y=0$),
are introduced to the Higgs sector in addition to
the original SM $\rm SU(2)_L$ doublet $\phi$ (with $\rm Y=1$).
The scalar content of the theory can be organized in terms of the $\rm SU(2)_L\otimes SU(2)_R$ symmetry.
In order to make this symmetry explicit, we write the doublet in the form of a bidoublet $\Phi$
and combine the triplets to form a bitriplet $\Delta$:
\begin{align}
\Phi=\left(
\begin{array}{cc}
\phi^{0*} & \phi^+ \\
\phi^-    & \phi^0
\end{array}
\right),\quad
\Delta=\left(
\begin{array}{ccc}
\chi^{0*} & \xi^+ & \chi^{++} \\
\chi^-    & \xi^0 & \chi^{+} \\
\chi^{--} & \xi^- & \chi^{0}
\end{array}
\right). \label{Higgs_matrices}
\end{align}
The custodial symmetry is preserved at tree level by imposing
such global $\rm SU(2)_L \otimes SU(2)_R$ symmetry on the scalar potential.
The Lagrangian can be obtained and V is the potential where the full formula can be found here:
\begin{align}
\begin{split}
\rm
V(\Phi, \, \Delta) \ =& \ \frac{1}{2} m_1^2 \, {\rm tr}[ \Phi^{\dagger} \Phi ] +
\frac{1}{2} m_2^2 \, {\rm tr}[ \Delta^{\dagger} \Delta ]
 +  \lambda_1 \left( {\rm tr}[ \Phi^{\dagger} \Phi ] \right)^2
 +  \lambda_2 \left( {\rm tr}[ \Delta^{\dagger} \Delta ] \right)^2
\\
&\rm
 +  \lambda_3 {\rm tr}\left[ \left( \Delta^{\dagger} \Delta \right)^2 \right]
 +  \lambda_4 {\rm tr}[ \Phi^{\dagger} \Phi ] {\rm tr}[ \Delta^{\dagger} \Delta ]
 +  \lambda_5 {\rm tr}\left[ \Phi^{\dagger} \frac{\sigma^a}{2} \Phi \frac{\sigma^b}{2} \right]
                  {\rm tr}[ \Delta^{\dagger} T^a \Delta T^b]
\\
&\rm
 + \mu_1 {\rm tr}\left[ \Phi^{\dagger} \frac{\sigma^a}{2} \Phi \frac{\sigma^b}{2} \right]
                               (P^{\dagger} \Delta P)_{ab}
 + \mu_2 {\rm tr}[ \Delta^{\dagger} T^a \Delta T^b]
                               (P^{\dagger} \Delta P)_{ab} ~.
\end{split}
\label{potential}
\end{align}
After the symmetry, the scalar fields in the GM model can be classified into different representations
under the custodial symmetry transformation. The scalar fields from doublet $\Phi$
is decomposed into a 3-plet and a singlet.
Those from the triplet $\Delta$ is decomposed into a 5-plet, a 3-plet and a singlet.
Among these $\rm SU(2)_C$ multiplets, the 5-plet states directly become physical Higgs bosons,
{\it i.e.}, $\rm H_5=(H_5^{\pm\pm}, H_5^\pm, H_5^0$).
For the two 3-plets, one of the linear combinations corresponds to physical Higgs field,
{\it i.e.}, $\rm H_3=(H_3^\pm,H_3^0$), and the other becomes the Nambu-Goldstone bosons
$\rm G^\pm$ and $\rm G^0$ which are absorbed into the
longitudinal components of the $\rm W^\pm$ and $\rm Z$ bosons, respectively.
The GM model has several features. It preserves the
$\rho$-parameter unity at the tree level via custodial symmetry.
It offers the possibility of implementing the seesaw mechanism
to endow the neutrinos with naturally light Majorana masses.
The tree level couplings of the SM-like Higgs to fermions
and vector bosons may be enhanced in comparison to the SM.
The appearance of the $\rm H^{\pm}W^{\mp}Z$ coupling at the tree level,
and the presence of doubly-charged scalar particles, are phenomenal quite interesting.

\section{The Calculation}

\begin{figure}[hbtp]
\centering
\includegraphics[scale=0.4]{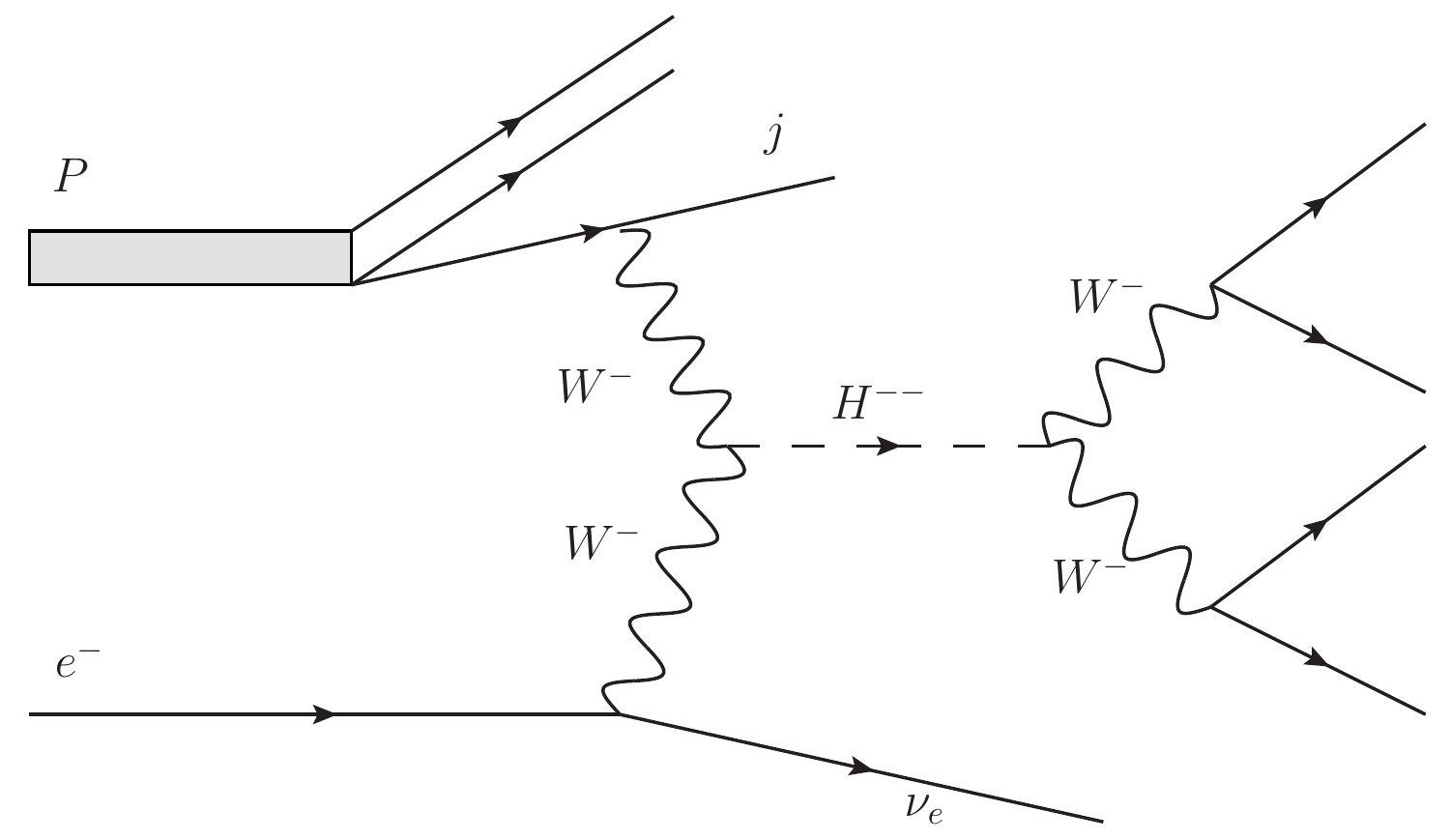}
\includegraphics[scale=0.2]{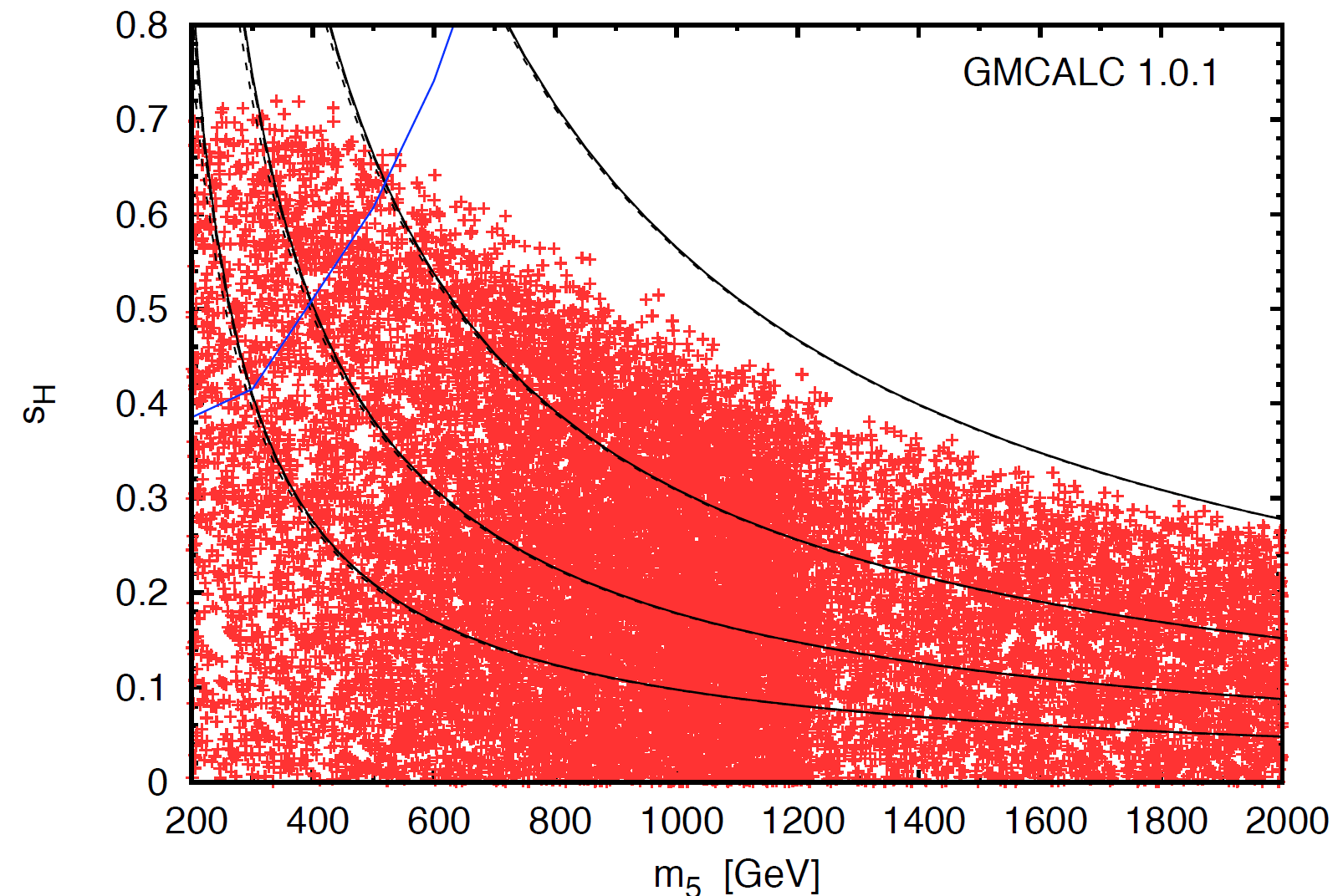}
\caption{\label{fig1_sig}
Left: Feynman diagram for the signal process $\rm e^- p \to \nu_e (H_5 ^{--}\to W^- W^- ) j $ with
$\rm \nu_e \bar\nu_\ell \bar\nu_\ell \ell^- \ell^- j$, $\rm \nu_{e} \bar\nu_\ell \ell^- jjj$ and $\rm jjjjj$ decay modes at the ep collider.
Right: The scatter point plot of the allowed parameter space in the plane of $\rm sin\theta_H$ and $\rm M_{H_5}$.}
\end{figure}

The fiveplet scalar states are produced primarily through the vector boson fusion (VBF), Drell-Yan
and associated production ($\rm VH_5$) modes. Our purpose is to study the prospects
for detecting the doubly-charged Higgs boson in the GM model at the ep colliders.
The proposed ep collider \cite{LHeC_1, LHeC_3, FCC-eh_2} is the Large Hadron Electron collider (LHeC),
which is a combination of 60 GeV electron beam and 7 TeV proton beam of the LHC.
This may later be extended to 50 TeV proton beam of Future Circular Collider (FCC-eh).
Both projects will create new electron facilities.
Through ep collision, the most promising channel is via the VBF mechanism.
We study three decay modes: one is the same-sign leptonic decay mode (mode A), the second is the
semileptonic decay mode (mode B) and the third is the multi-jet decay mode (mode C):
\begin{eqnarray}\nonumber
&&\rm mode\ A:\ \ e^-p \to \nu_{e} ( H_5^{--} \to W^-W^-) j \to \nu_{e} \bar\nu_\ell \bar\nu_\ell \ell^- \ell^- j \\\nonumber
&&\rm mode\ B:\ \ e^-p \to \nu_{e} ( H_5^{--} \to W^-W^-) j \to \nu_{e} \bar\nu_\ell \ell^- jjj \\
&&\rm mode\ C:\ \ e^-p \to \nu_{e} ( H_5^{--} \to W^-W^-) j \to \nu_{e} jjjjj,
\end{eqnarray}
where j refers to jets with $\rm j=u,c,\bar{d},\bar{s}$ and other possibilities from W boson decay, $\rm \ell=e,\mu$ are the leptons.
The corresponding feynman diagram is plotted in Fig.\ref{fig1_sig}[left].
The large multi-jet backgrounds make modes B and C the challenge choices
to detect the doubly-charged Higgs bosons.
For mode A, by distinguishing $\rm e,\ \mu$ leptons,
we can have $\rm e^-e^-$, $\rm e^-\mu^-$ or $\rm \mu^-\mu^-$ final combinations of the observables.
Since electron is the initial collision particle, it's easier to appear in the final state.
For $\rm e^-e^-$ and $\rm e^-\mu^-$ combinations, they will be affected
by the $\rm e^- p \to e^- W^- j$ background, which is unfortunately quite large.
In contrast, $\rm \mu^-\mu^-$ final state will have much clean environment and become the advantaged one.
The scan over the parameter space is performed in Fig.\ref{fig1_sig}[right]
(figure taken from Ref.\cite{LHC_H5}). The two important parameters are $\rm M_{H_5}$
and $\rm \sin\theta_H$ since our studied coupling
and production rates are directly proportional to or depend on them.
The area above the blue curve is excluded by the LHC experimental data (update in Ref.\cite{H5pplimit}).
Considering the doubly-charged Higgs decay, when $\rm M_{H_5}$ is not very large,
for over $98\%$ of the scan points, $\rm BR(H^{\pm\pm}_5 \to W^\pm W^\pm)$ were above $99\%$.
Therefore, for simplicity, the $\rm H^{\pm\pm}_5$ state can be assumed to decay entirely into vector boson
pairs for masses above the same-sign W pair threshold, say, $\rm BR(H^{\pm\pm}_5 \to W^\pm W^\pm)=1$.
\begin{figure}[hbtp]
\centering
\includegraphics[scale=0.5]{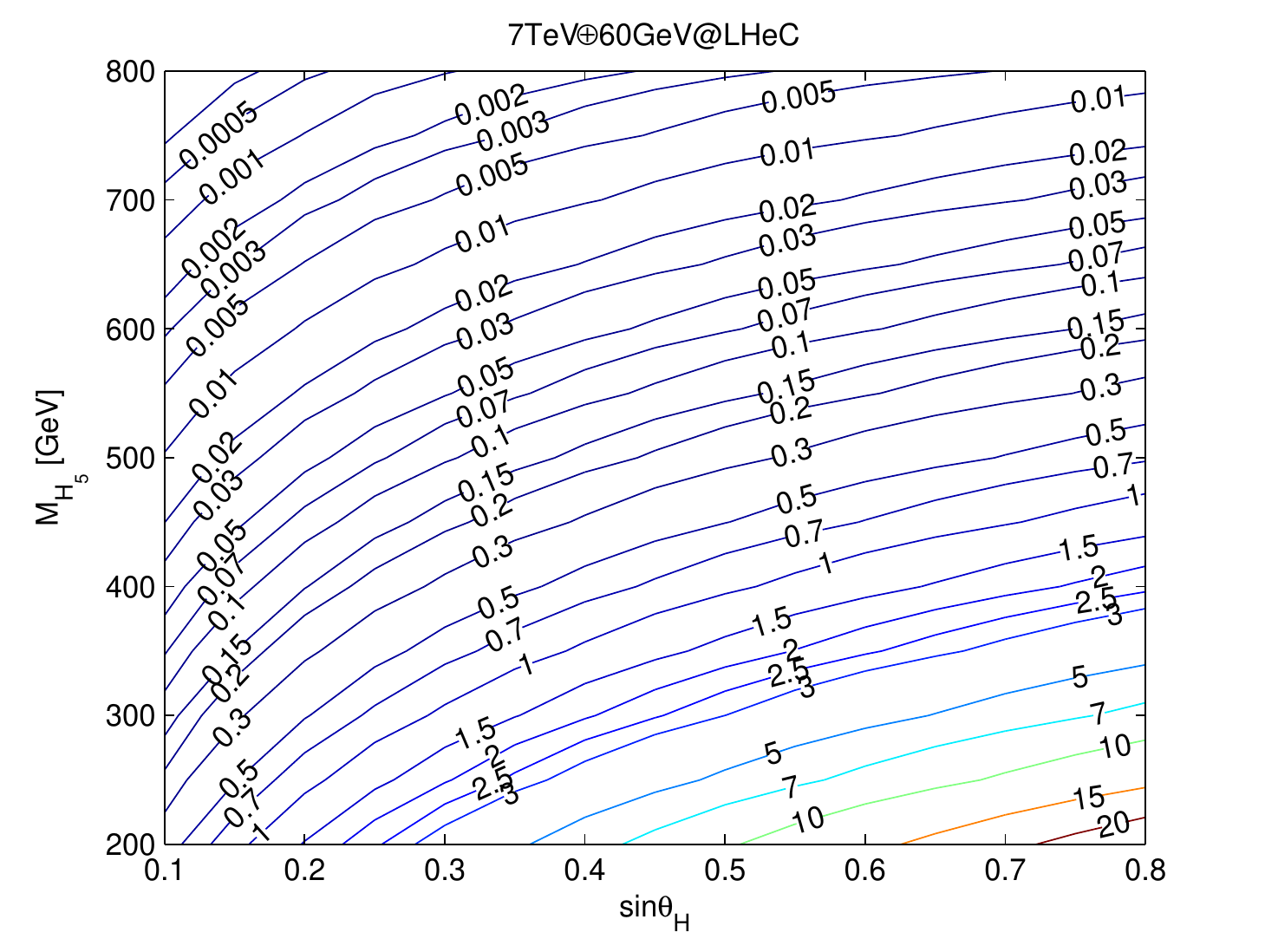}
\includegraphics[scale=0.5]{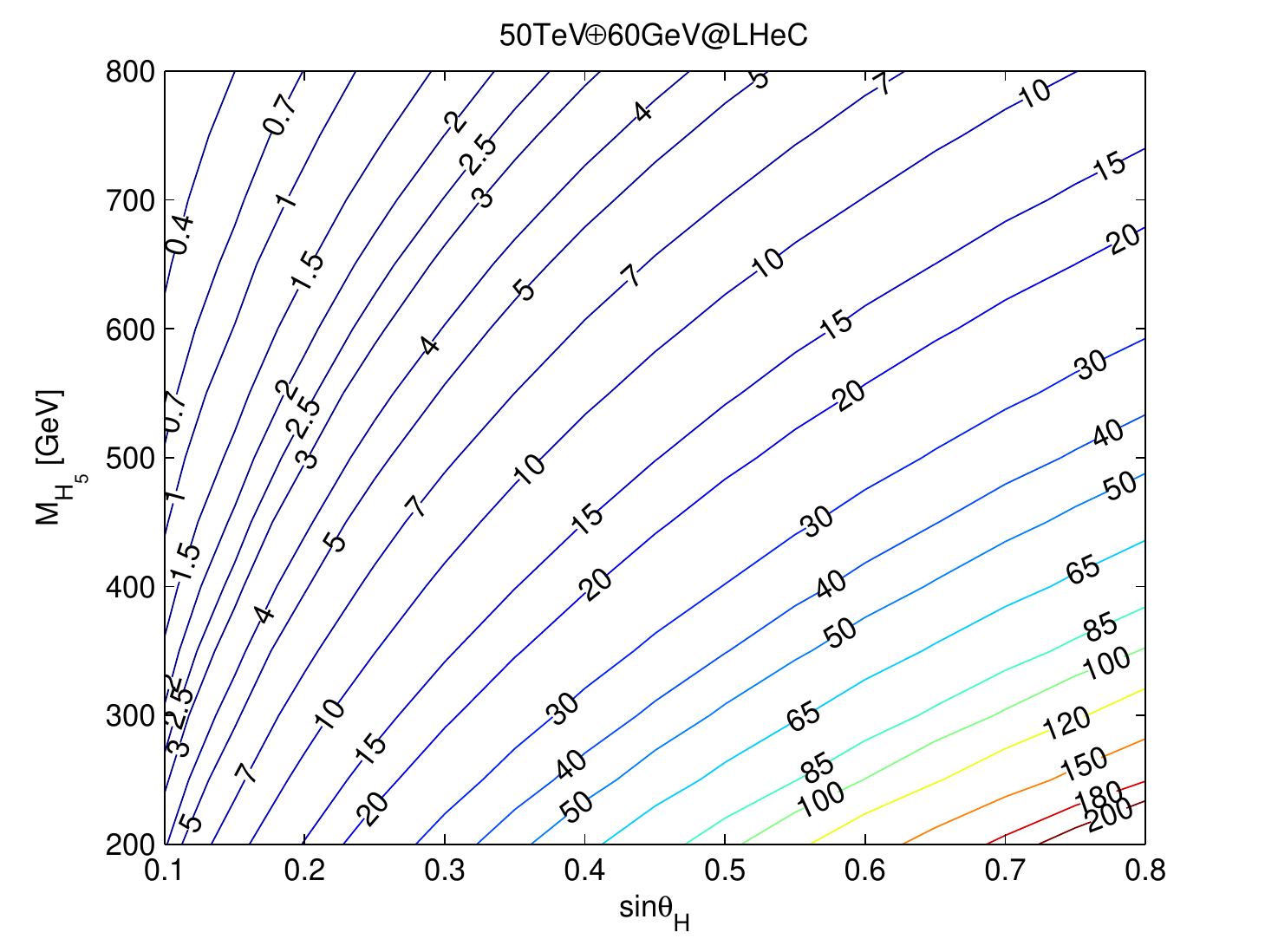}
\caption{\label{GM parameter scan}
Total cross sections (in $\rm fb$) for 5-plet Higgs production
$\rm e^-p\to\nu_e H_5^{--}j$ as the functions of $\rm M_{H_5}$ and $\rm sin\theta_H $.
(a),(b) correspond to the 7 TeV LHeC and 50 TeV FCC-eh, respectively. The electron beam is 60 GeV.}
\end{figure}
Before doing the full signal and background simulation,
we present a general VBF production rate of the doubly-charged Higgs boson
at the LHeC and FCC-eh in Fig.\ref{GM parameter scan}.
The total cross sections are plotted in fb as functions of $\rm sin\theta_H$ and $\rm M_{H_5}$.
The allowed boundary in Fig.\ref{fig1_sig} should be taken into account.
The cross section at the LHeC is around the order of only 0.1-1 fb,
while at the FCC-eh, it can be much larger. We consider the full simulation chain including the detector effects
at both the LHeC and the FCC-eh, thanks to the contributions from the LHeC working group.
\begin{figure}[hbtp]
\centering
\includegraphics[scale=0.25]{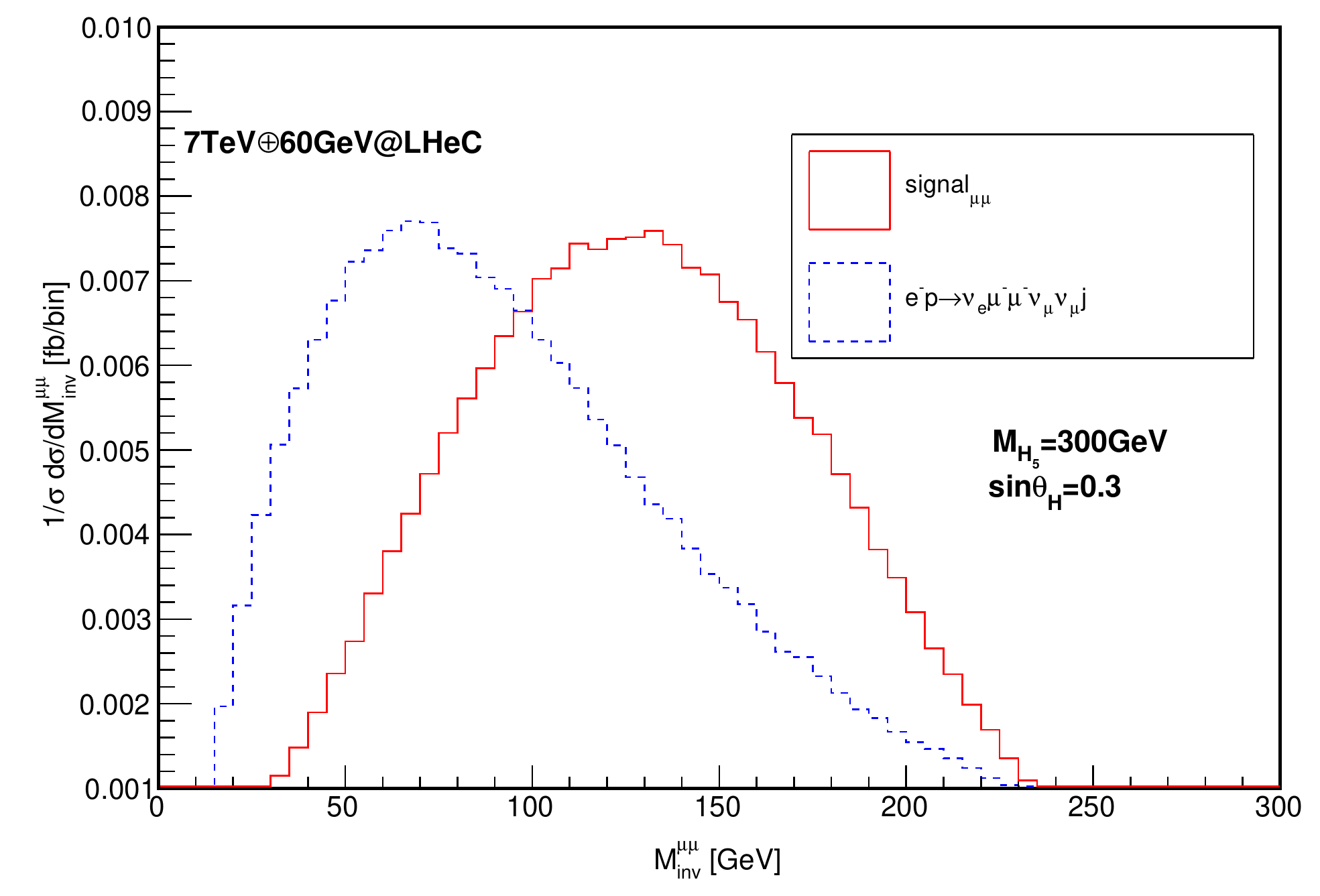}
\includegraphics[scale=0.25]{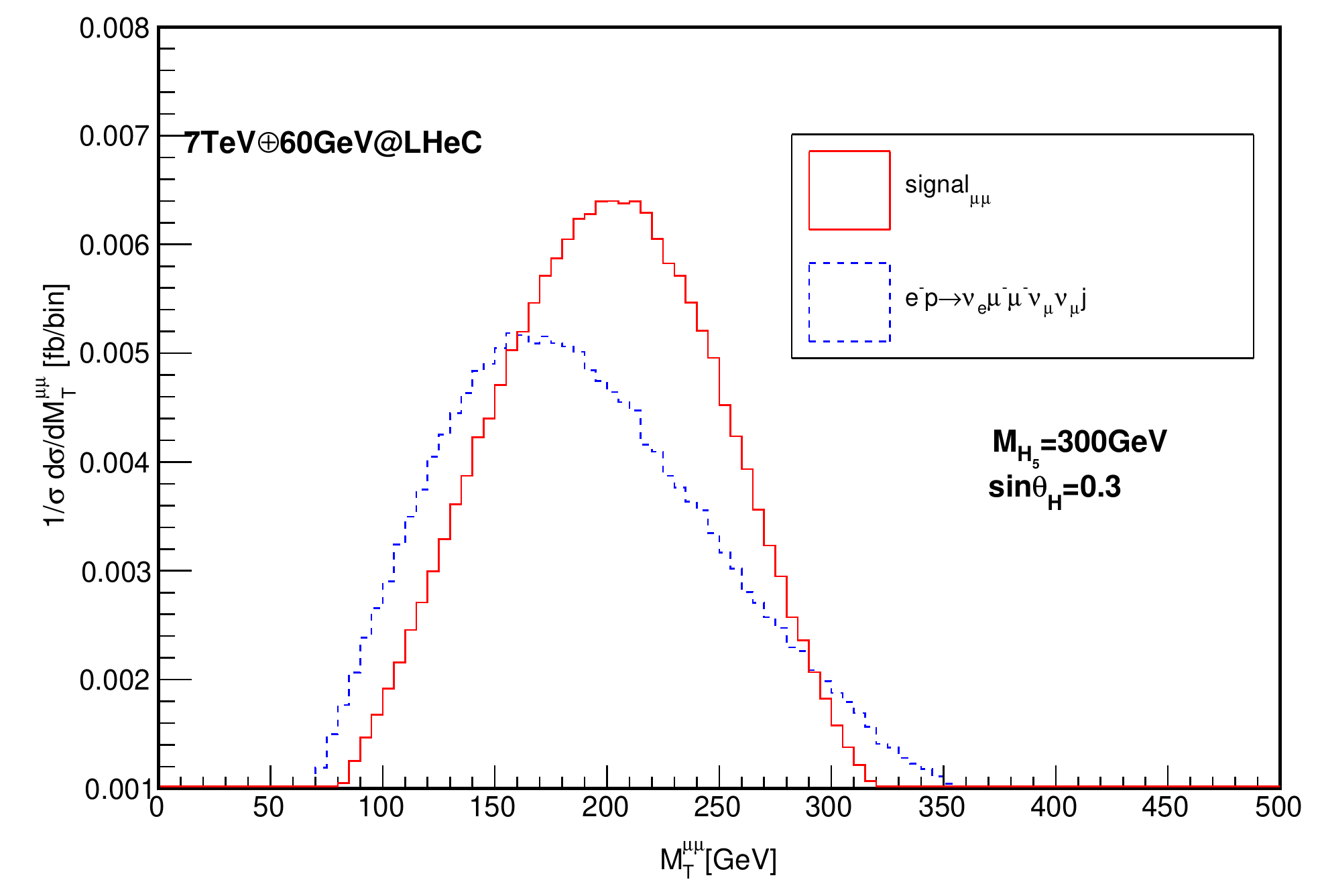}\\
\includegraphics[scale=0.25]{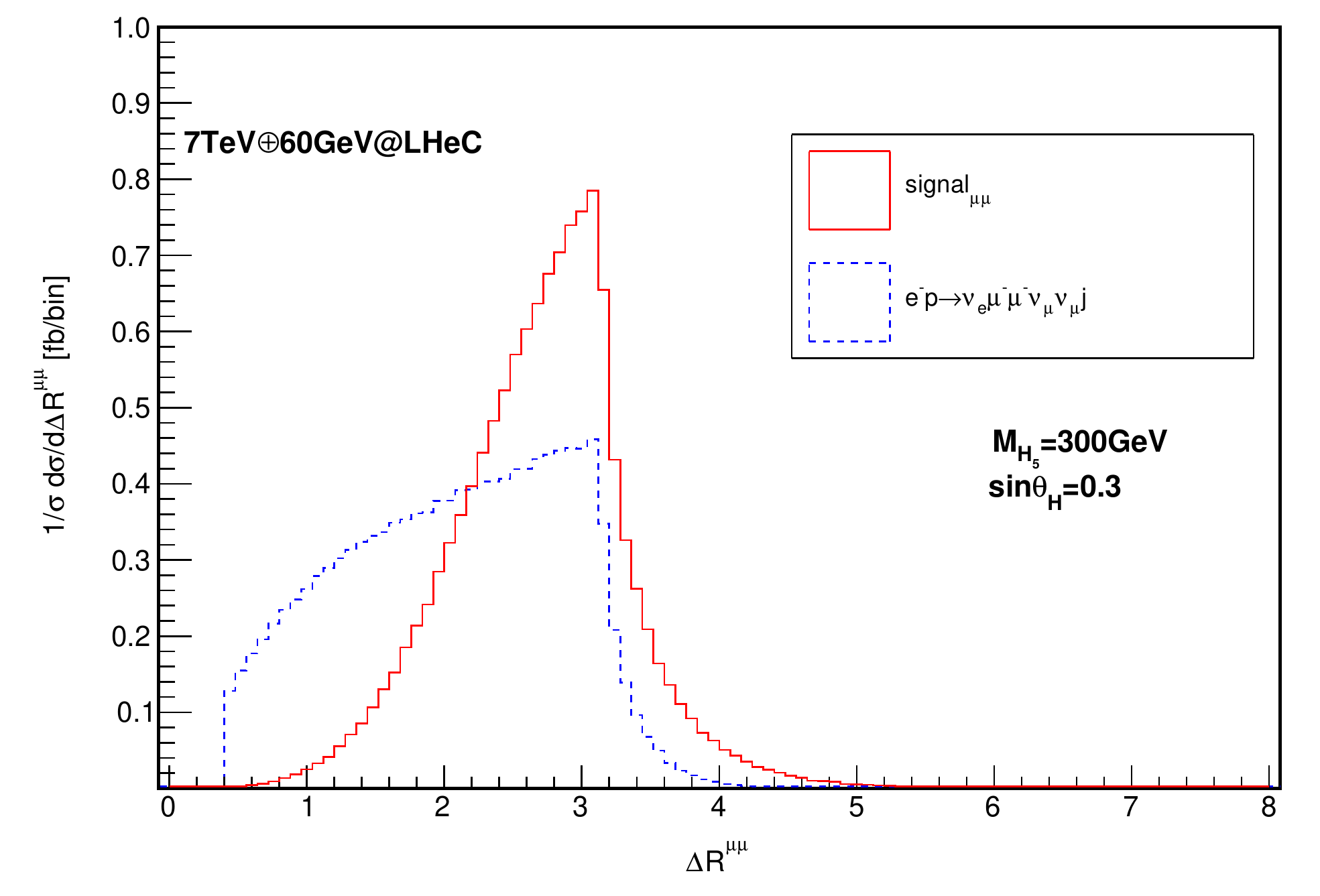}
\includegraphics[scale=0.25]{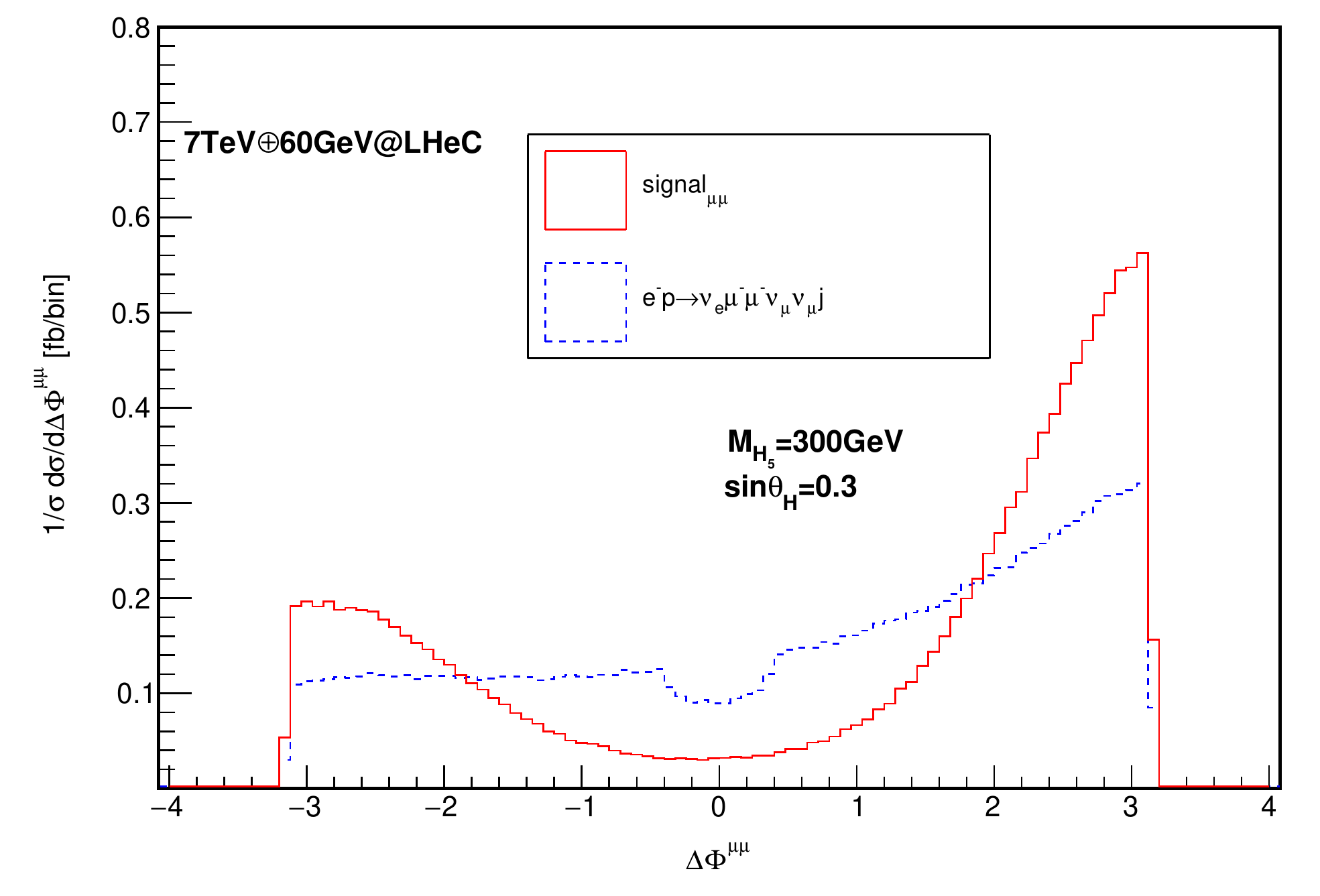}
\caption{\label{7_distributions}
Various kinematical distributions (in units of fb per bin)
for the signal and backgrounds at the 7 TeV LHeC. The electron beam is 60 GeV.
Here $\rm M_{H_5}=300\ GeV$ and $\rm \sin\theta_H=0.3$. Plots are unit normalized.}
\end{figure}
The studied topology is characterized by a same-sign $\mu^-\mu^-$ lepton pair,
a forward rapidity jet plus missing transverse momentum from the undetected neutrinos.
The dominant background is simply $\rm e^-p\to\nu_e \mu^-\mu^-\nu_\mu\nu_\mu j$,
including 0-$\rm W^-$, 1-$\rm W^-$ and 2-$\rm W^-$ contributions, where 2-$\rm W^-$
($\rm e^-p\to\nu_e (W^-W^-\to \mu^-\nu_\mu\mu^-\nu_\mu)j$) is the dominant one.
In Fig.\ref{7_distributions} we present various kinematical distributions
for both the signal and background. Results are for 7 TeV LHeC
while for 50 TeV FCC-eh we can get similar ones.
\begin{table}[htbp]
\begin{center}
%\begin{ruledtabular}
\begin{tabular}{c | c  c  c  c  c  }
\hline
\makecell {7TeV$\oplus$60GeV$@$LHeC \\ $[\rm 1\ ab^{-1}]$ } & \makecell{ same-sign \\ $\mu^-\mu^-$}
& \makecell { $\rm \Delta R^{\mu\mu}$ \\ > 1.92 }  & \makecell{ $\rm \Delta\Phi^{\mu\mu}\in (-\pi, -1.28)$\\ or\ $(1.36,\pi)$ }
&  \makecell{ $\rm M_{inv}^{\mu\mu}$\\ >75\ GeV }   &   \makecell{  $\rm M_{T}^{\mu\mu}$ \\ >40\ GeV  } \\
\hline
$\rm signal_{\mu\mu}$[$\rm \sin\theta_{H}=0.3$]         &  10.46  & 9.45   & 9.24  & 9.21  &  10.23    \\
\hline
$\rm B$[$\rm e^-p\to\nu_e \mu^-\mu^-\nu_\mu\nu_\mu j$]  & 10.52    & 6.43  & 7.13  & 6.71  & 9.58  \\
\hline
$SS$                                            &  2.84    & 3.13  & 2.96   & 3.02  & 2.89  \\
\hline\hline
\makecell {50TeV$\oplus$60GeV$@$FCC-eh \\ $[\rm 100\ fb^{-1}]$ }  & \makecell{ same-sign \\ $\mu^-\mu^-$}
& \makecell { $\rm \Delta R^{\mu\mu}$ \\ > 1.6 }  & \makecell{ $\rm \Delta\Phi^{\mu\mu}\in (-\pi, -0.88)$\\ or\ $(1.04,\pi)$ }
&  \makecell{ $\rm M_{inv}^{\mu\mu}$\\ >70\ GeV }   &   \makecell{  $\rm M_{T}^{\mu\mu}$ \\ >78\ GeV  } \\
\hline
$\rm signal_{\mu\mu}$[$\rm \sin\theta_{H}=0.3$]&  10.16  & 9.44 & 9.25  &  9.13 &   9.75   \\
\hline
$\rm B$[$\rm e^-p\to\nu_e \mu^-\mu^-\nu_\mu\nu_\mu j$]  & 8.29  & 5.88  & 6.41  & 5.40  & 7.04  \\
\hline
$SS$                  &  3.03  & 3.23 &  3.08  & 3.24  & 3.11  \\
\hline
\end{tabular}
%\end{ruledtabular}
\caption{\label{SBaftercuts}
Expected number of events and signal significance ($SS$)
evaluated with $\rm 1\ ab^{-1}$ integrated luminosity at the LHeC and  $\rm 100\ fb^{-1}$ at the FCC-eh.
Here $\rm M_{H_5}=300\ GeV$ and each time take only one cut.}
\end{center}
\end{table}
We will aim at finding the most efficient selections in order to allow the
best separation between noise-related and signal-related events.
We compare the efficiency of different cuts in Tab.\ref{SBaftercuts}
include $\rm M_{inv}$, $\Delta \Phi$, $\rm \Delta R$ and $\rm M_T$
for the $\mu^-\mu^-$ system for both the 1 $\rm ab^{-1}$ LHeC and 100 $\rm fb^{-1}$ FCC-eh.
Each time we take only one cut, so as not cut events too much, otherwise the significance can be reduced.
For $\rm M_{H_{5}}$ not very large, $\rm \Delta R$ is the one we use that can lead better
significance ($SS\rm = \sqrt{2[(n+b)\log(1+n/b)-n]}$,
where n is the number of events and b is the number of backgrounds respectively).
\begin{figure}[hbtp]
\centering
\includegraphics[scale=0.35]{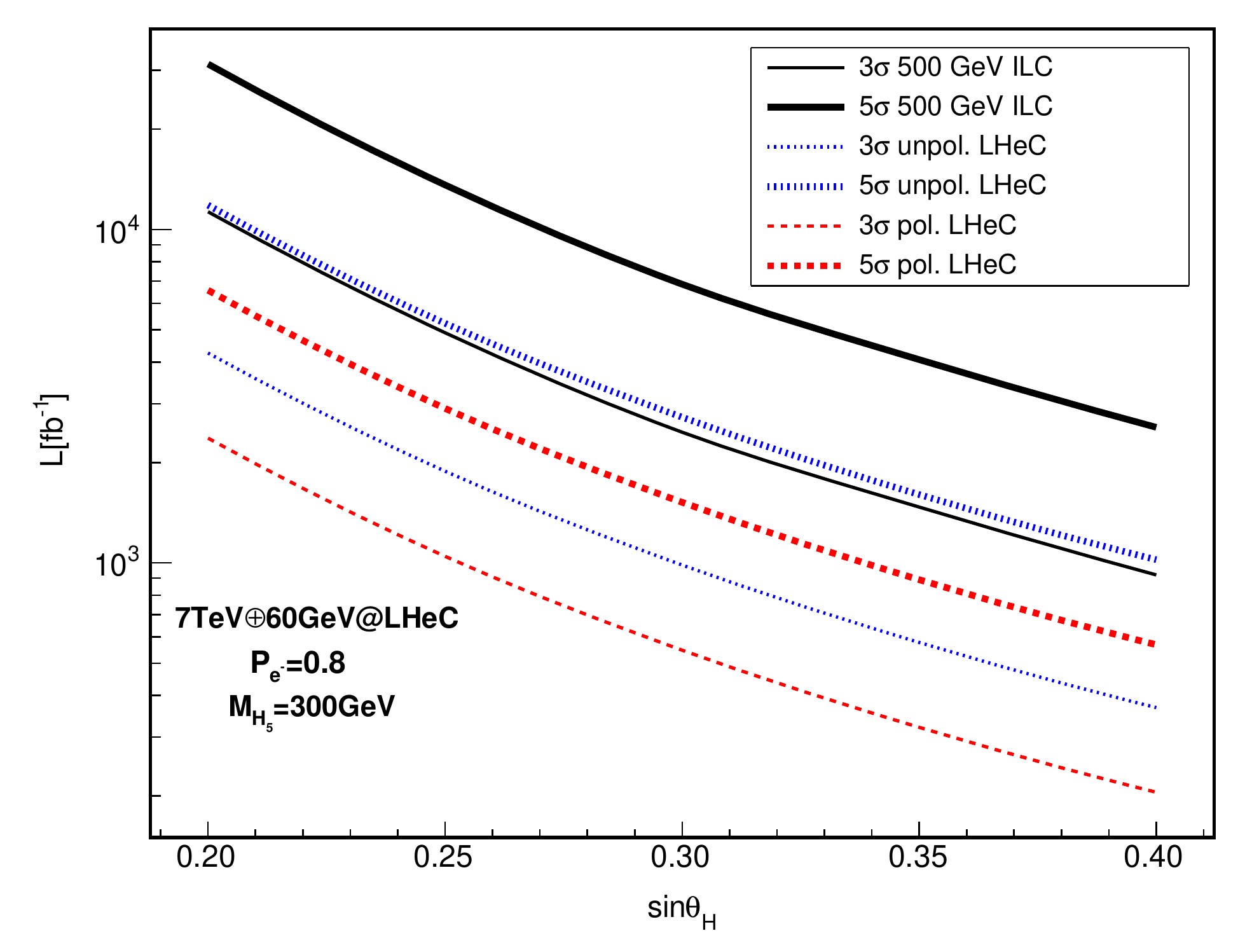}
\includegraphics[scale=0.35]{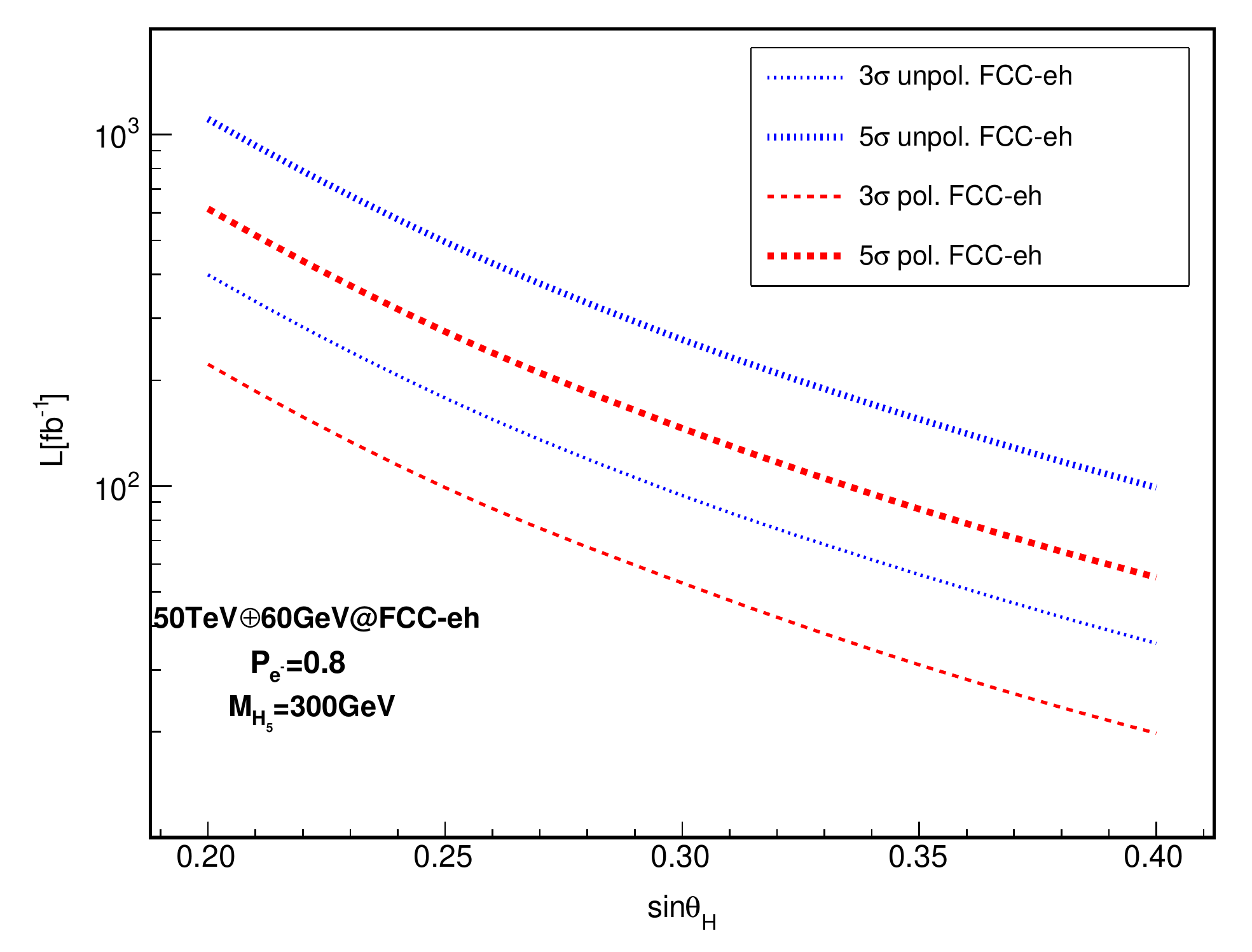}
\vspace{-0.3cm}
\caption{\label{luminosity}
The lowest necessary luminosity with $3\sigma$ and $5\sigma$
discovery significance as a function of $\rm sin\theta_H$.
Here we used $10\%$ systematic uncertainty for background yields only.}
\end{figure}
In Fig.\ref{luminosity} we calculate the lowest necessary luminosity with $3\sigma$
and $5\sigma$ discovery significance as a function of $\rm sin\theta_H$.
We present the results for the LHeC and FCC-eh
with both the unpolarized and polarized ($\rm p_{e^-}=0.8$) electron beams.
We used $10\%$ systematic uncertainty for background yields only.
It is shown that high integrated luminosity is required to probe small $\rm sin\theta_H$.
At the FCC-eh, the lowest necessary luminosity are much reduced than that at the LHeC as expected.
Results in comparison with others from a related study
at the International Linear Collider are also shown in \cite{ILC_sinh}.
We see that the lowest necessary luminosity are much smaller.
In addition, polarized beam could make the measurement even better.

\section{Summary}

The possible existence of heavy exotic particles in beyond Standard Models are highly expected.
The Georgi-Machacek (GM) model is one of such scenarios with an extended scalar sector
which can group under the custodial $\rm SU(2)_C$ symmetry.
There are 5-plet, 3-plet and singlet Higgs bosons under the classification
of such symmetry in addition to the SM Higgs boson.
In the GM model, there are 5-plets doubly-charged states
so that the distinct phenomenological features should appear.
In this talk, we present the 5-plet Higgs production at the ep colliders.
We focus on the vector boson fusion channel
and decays into final states containing a pair of same-sign $\mu^-$.
The discovery significance are calculated as the functions of the triplet vacuum expectation value
and the lowest necessary luminosity. We expect the future ep experiments
could measure the strength of the doubly charged Higgs boson production
or otherwise put stringent constraints on it.


\begin{thebibliography}{99}

\bibitem{LHeC_1}
M. Klein, \emph{The Large Hadron Electron Collider Project}, \emph{Proceedings,
17th International Workshop on Deep-Inelastic Scattering and Related Subjects (DIS 2009),
Madrid, Spain}, April 26-30, 2009, {\tt[arXiv:0908.2877]}.

\bibitem{LHeC_3}
O. Bruening and M. Klein, \emph{The Large Hadron Electron Collider},
\emph{Mod. Phys. Lett. A} 28 (2013) no.16, 1330011, {\tt [arXiv:1305.2090]}.

\bibitem{FCC-eh_2}
M. Klein, \emph{Deep inelastic scattering at the energy frontier}, \emph{Annalen Phys}. 528 (2016) 138-144.

\bibitem{LHC_H5}
Marco Zaro and Heather Logan, \emph{Recommendations for the interpretation of LHC searches for
$H^0_5$, $H^\pm_5$, and $H^\pm_5$ in vector boson fusion with decays to vector boson pairs},
\emph{LHCHXSWG-2015-001}.

\bibitem{H5pplimit}
[CMS Collaboration], \emph{Observation of electroweak production of same-sign W boson pairs
in the two jet and two same-sign lepton final state in pp collisions at 13 TeV}, \emph{CMS-PAS-SMP-17-004}.

\bibitem{ILC_sinh}
YuFei Zhang, Hao Sun, Xuan Luo and WeiNing Zhang,
\emph{Searching for the heavy charged custodial fiveplet Higgs boson in the Georgi-Machacek model at the
International Linear Collider}, \emph{Phys.Rev.D} {\bf 95} (2017) 115022, [{\tt arXiv:1706.01490}].

\end{thebibliography}
\end{document}